\documentclass[sigconf, screen]{acmart}
\usepackage[capitalise]{cleveref}
\usepackage{multirow}
\usepackage{algorithm}
\usepackage{algpseudocode}
\usepackage{hyperref}
\usepackage{newtxtext,newtxmath}
\usepackage{amsmath}
\usepackage{calligra}
\usepackage{amsfonts}

\AtBeginDocument{%
  }

\setcopyright{none}
\renewcommand\footnotetextcopyrightpermission[1]{}
\acmConference[ASPDAC '25]{30th Asia and South Pacific Design Automation
Conference}{January 20--23, 2025}{Tokyo, Japan}
\begin{document}

% \title{PI\textit{Mutation}: Exploring the Potential of Real PIM Architecture \\ for Quantum Circuit Simulation}

\title{PI\textit{Mutation}: Exploring the Potential of PIM Architecture for Quantum Circuit Simulation}

%\email{webmaster@marysville-ohio.com}
%\affiliation{%
%  \institution{Institute for Clarity in Documentation}
%  \streetaddress{P.O. Box 1212}
%  \country{USA}

\author{Dongin Lee}
\authornote{This work was conducted while he was affiliated with Yonsei University.}
\authornote{These two authors contributed equally to this work.}
\affiliation{
  \institution{National University of Singapore}
  \city{Singapore}
  \country{Singapore}
}
\email{dongin.lee@u.nus.edu}
\orcid{0009-0000-5624-5494}

\author{Enhyeok Jang}
\authornotemark[2]
\affiliation{
  \institution{Yonsei University}
  \city{Seoul}
  \country{Korea}
}
\email{enhyeok.jang@yonsei.ac.kr}
\orcid{0009-0000-7034-6793}

\author{Seungwoo Choi}
\affiliation{
  \institution{Yonsei University}
  \city{Seoul}
  \country{Korea}
}
\email{seungwoo.choi@yonsei.ac.kr}
\orcid{0009-0005-2162-8993}

\author{Junwoong An}
\affiliation{
  \institution{Yonsei University}
  \city{Seoul}
  \country{Korea}
}
\email{junwoong.an@yonsei.ac.kr}
\orcid{0009-0007-5674-3099}

\author{Cheolhwan Kim}
\affiliation{
  \institution{Yonsei University}
  \city{Seoul}
  \country{Korea}
}
\email{cheolhwan.kim@yonsei.ac.kr}
\orcid{0009-0005-1307-4070}

\author{Won Woo Ro}
\affiliation{
  \institution{Yonsei University}
  \city{Seoul}
  \country{Korea}
}
\email{wro@yonsei.ac.kr}
\orcid{0000-0001-5390-6445}

\renewcommand{\shortauthors}{Dongin Lee, Enhyeok Jang et al.}

\begin{abstract}

% Quantum circuit simulation, which utilizes the power of classical computation, is crucial for verifying quantum algorithms, mainly because currently accessible quantum computing devices are slow and prone to errors. 
Quantum circuit simulations are essential for the verification of quantum algorithms on behalf of real quantum devices.
% However, as the number of qubits increases, the memory requirements for such simulations grow exponentially.
However, the memory requirements for such simulations grow exponentially with the number of qubits involved in quantum programs. 
Moreover, a substantial number of computations in quantum circuit simulations cause low locality data accesses, as they require extensive computations across the entire table of the full state vector. 
These characteristics lead to significant latency and energy overheads during data transfers between the CPU and main memory. 
Processing-in-Memory (PIM), which integrates computational logic near DRAM banks, could present a promising solution to address these challenges.

% In this paper, we introduce PI\textit{Mutation} (PIM framework for qUanTum circuit simulATION), the first attempt to leverage UPMEM, which is a publicly available PIM-integrated DIMM, for achieving the fast and low-power quantum circuit simulation.
% To mitigate the overhead associated with quantum circuit simulation due to the hardware constraints in the real UPMEM system, PI\textit{Mutation} incorporates three optimization strategies: (i) gate merging, (ii) row swapping, and (iii) vector partitioning. 
% Our evaluations show that PI\textit{Mutation} achieves an average speedup of 2.99$\times$ and 16.51$\times$ while reducing energy consumption by 25.23\% and 75.29\%, over QuEST simulator on CPU for the 16 and 32-qubit benchmarks, respectively. 
In this paper, we introduce PI\textit{Mutation} (PIM framework for qUanTum circuit simulATION) for achieving fast and energy-efficient quantum circuit simulation.
PI\textit{Mutation} is the first attempt to leverage UPMEM, a publicly available PIM-integrated DIMM, to implement quantum circuit simulations. 
PI\textit{Mutation} incorporates three optimization strategies to overcome the overhead of quantum circuit simulation using the real PIM system: (i) gate merging, (ii) row swapping, and (iii) vector partitioning.
Our evaluations show that PI\textit{Mutation} achieves an average speedup of 2.99$\times$ and 16.51$\times$ with a reduction of energy of 25.23\% and 75.29\% over the QuEST simulator on CPU in 16- and 32-qubit benchmarks, respectively.

% Processing-in-Memory (PIM), which leverages computational logic near DRAM banks, can efficiently handle quantum circuit simulations.
% Processing-in-Memory (PIM), which leverages computational logic near DRAM banks to efficiently handle memory-bounded workloads, can efficiently handle quantum circuit simulations.

% In our experiments, for a single DPU,PI\textit{Mutation} reduces run-time by 86.2\% and reduces energy consumption by 14.5\% compared to the naive state vector simulation on UPMEM.
% For multiple DPUs,PI\textit{Mutation} shows 147\% faster simulation speeds with 4 DPUs and 88\% faster with 8 DPUs compared to the QuEST simulator on CPU.

\end{abstract}

\begin{CCSXML}
<ccs2012>
   <concept>
       <concept_id>10010583.10010786.10010787.10010788</concept_id>
       <concept_desc>Hardware~Emerging architectures</concept_desc>
       <concept_significance>500</concept_significance>
       </concept>
   <concept>
       <concept_id>10010583.10010786.10010787.10010790</concept_id>
       <concept_desc>Hardware~Emerging simulation</concept_desc>
       <concept_significance>300</concept_significance>
       </concept>
   <concept>
       <concept_id>10010520.10010521.10010528</concept_id>
       <concept_desc>Computer systems organization~Parallel architectures</concept_desc>
       <concept_significance>500</concept_significance>
       </concept>
   <concept>
       <concept_id>10010520.10010521.10010542.10010546</concept_id>
       <concept_desc>Computer systems organization~Heterogeneous (hybrid) systems</concept_desc>
       <concept_significance>500</concept_significance>
       </concept>
 </ccs2012>
\end{CCSXML}

\ccsdesc[500]{Hardware~Emerging architectures}
\ccsdesc[300]{Hardware~Emerging simulation}
\ccsdesc[500]{Computer systems organization~Parallel architectures}
\ccsdesc[500]{Computer systems organization~Heterogeneous (hybrid) systems}

\keywords{Near-Data Processing, Processing-in-Memory, State Vector-Based Quantum Circuit Simulation}


\maketitle

\section{Introduction}

Quantum circuit simulations serve as a crucial tool for validating and evaluating quantum algorithms, particularly in small qubit systems, compensating for the limitations of currently available inaccurate quantum computing devices. 
Unlike quantum computing devices, where measurements inherently collapse quantum superposition states, quantum circuit simulations allow for the investigation of intermediate processes in quantum state evolution, enabling the verification of errors \cite{li}. 
However, these simulations face significant challenges, particularly due to data access of low locality during quantum gate operations and the exponential growth in memory capacity requirements as the number of qubits increases.

Processing-in-Memory (PIM) technology, which integrates computing capabilities directly inside or in close proximity to memory chips, has the potential to reduce energy consumption and latency. 
This could be achieved through increased internal bandwidth and reduced physical distance between processing units and memory \cite{gomez}. 
Previous studies have successfully applied PIM architectures to various domains, including graph processing, genome analysis, deep neural network (DNN) acceleration, and embedding tables for recommendation systems \cite{ahn2015scalable, gupta2019rapid, lee2021hardware, kal}, leveraging the inherent advantages of PIM. 
Despite this, the application of PIM systems to quantum circuit simulations remains largely unexplored. 
Our analysis of quantum circuit simulation workloads reveals that a significant portion of benchmarks exhibit memory-bounded characteristics, suggesting that PIM could offer a viable solution for addressing the challenges of quantum circuit simulation.

In this work, we propose PI\textit{Mutation} (\underline{PIM} framework for q\underline{U}an\underline{T}um circuit simul\underline{ATION}, which implements a PIM framework specifically designed for quantum circuit simulation. 
To the best of our knowledge, our work introduces the first framework for quantum circuit simulation utilizing UPMEM DIMMs, the first commercially available DRAM-based PIM architecture \cite{devaux}. 
It is important to note that there are discrepancies between real PIM architectures and PIM architectures evaluated by simulators or prototypes in numerous studies.
Specifically, the real PIM architecture presents certain design considerations that should be accounted for when developing an efficient quantum circuit simulation framework.

Firstly, the DRAM Processing Unit (DPU), which is the processing unit within UPMEM, only supports integer operations. 
Secondly, the UPMEM system does not support direct communication capabilities between DPUs. 
Therefore, it is crucial to consider these characteristics of the UPMEM PIM system when implementing the framework. 
To optimize the performance of quantum circuit simulation while accommodating the characteristics of real PIM architecture, PI\textit{Mutation} incorporates three key optimization techniques, as outlined below.

\textbf{Gate Merging to Exploit UPMEM Native Integer Operations}:
PI\textit{Mutation} optimizes quantum circuit simulations by replacing floating-point division operations with addition, subtraction, and bit-wise shift operations, which are natively supported on UPMEM hardware. 
This is achieved by merging single-qubit gates, such as H, RX($\pi/2$), and RY($\pi/2$), into even numbers. 
The core principle of this optimization lies in combining multiple quantum gates that have coefficients of irrational numbers to produce gates of integer-type coefficients, thereby enhancing compatibility with UPMEM hardware's processing capabilities.

\textbf{Row Swapping Instead of Matrix Multiplication}:
We observe that the matrix coefficients associated with gates such as X, CNOT, Toffoli, and SWAP are only 0 or 1. 
These gate operations effectively swap the values of state vectors. 
Consequently, these gates are implemented by directly swapping the positions of the coefficients involved in the operation, bypassing the need for matrix multiplication and thus simplifying the computational process.

\textbf{Vector Partitioning for Separable Quantum States}:
In cases where a quantum program exhibits separability, PI\textit{Mutation} partitions the state vector into separate vectors for each independent state and assigns these to different DPUs, which are UPMEM processing units. 
This approach treats the quantum states individually rather than as a whole state vector. 
After processing, the individual state vectors are reconstructed into the complete state vector on the Host CPU. 
This method can reduce the need for DPU communication and minimize the size of the state vectors, particularly in quantum simulation workloads where the separability condition is met.

We first evaluate the performance of Gate Merging and Row Swapping optimizations under a single DPU condition. 
The Gate Merging and Row Swapping techniques achieve speedups of 1.26$\times$ and 1.13$\times$, respectively, over a naive quantum state-vector simulation. 
We also assessed the performance of Vector Partitioning in combination with Gate Merging and Row Swapping by utilizing multiple DPUs. In this experiment, we compared PI\textit{Mutation} against QuEST \cite{quest}, a high-performance quantum circuit simulator based on a C-library.
PI\textit{Mutation} demonstrates speedups of 2.99$\times$ and 16.51$\times$ while achieving energy reductions of 25.23\% and 75.29\% compared to the QuEST simulator on a CPU for 16- and 32-qubit benchmarks, respectively.

The main contributions of this work are as follows:
\begin{itemize}
    \item We hypothesize that quantum circuit simulations can be executed efficiently on processing-in-memory platforms due to their memory and capacity-bound workload characteristics. 
    We test this hypothesis by using UPMEM, the first commercial Processing-in-Memory architecture.
    \item We observe that the real UPMEM architecture has some characteristics that must be considered when implementing an efficient quantum circuit simulation framework: 
    (i) UPMEM DPUs support only integer operations, and (ii) there is no direct communication channel between DPUs.
    \item We propose PI\textit{Mutation}, the first framework for quantum circuit simulation utilizing UPMEM DIMMs, which integrates three key optimization strategies: 
    (i) Gate Merging, (ii) Row Swapping, and (iii) Vector Partitioning.
    \item We evaluate the performance of PI\textit{Mutation} using 16- and 32-qubit benchmarks and demonstrate that PI\textit{Mutation} achieves average speedups of 2.99$\times$ and 16.51$\times$ while reducing energy consumption by 25.23\% and 75.29\%, respectively, compared to the QuEST simulator on a CPU.
\end{itemize}

\section{Preliminaries}

This section explains the foundational concepts of quantum computing, including the principles of qubit superposition and entanglement, as well as their representation using state vectors.
Furthermore, it discusses the implementation of quantum state vector simulation, a commonly employed method for simulating quantum circuits, and the resource requirements for simulating large qubit systems. 
The UPMEM architecture is then introduced, describing its processing elements and memory structure, which offer opportunities for alleviating memory bandwidth constraints in quantum simulations. 

\subsection{Quantum Computing Basic}

Quantum computation relies on qubits, which differ from classical bits.
A quantum state \begin{math}|\psi\rangle\end{math} is represented by the combination of the basis vectors \begin{math}|0\rangle\end{math} and \begin{math}|1\rangle\end{math} by existing in a superposition of 0 and 1 states.

\begin{equation}\label{eq1}
    |\psi\rangle = a_0|0\rangle + a_1|1\rangle,
\end{equation}
where two possible states of a single qubit (\textit{i.e.,} 0 and 1 states) are expressed by two basis vectors.
Superposition allows us to describe qubits with complex probability amplitudes, denoted as $a_0$ and $a_1$ in \cref{eq1}. The squared amplitudes represent the measurement probability, and the sum of the amplitude is equal to as \cref{eq2}.

\begin{equation}\label{eq2}
    {a_0}^{2} + {a_1}^{2} = 1,
\end{equation}
For instance, a qubit with amplitudes of \begin{math}\frac{1}{2}\end{math} and \begin{math}\frac{\sqrt{3}}{2}\end{math} has a 0.25 probability of being measured as 0 and a 0.75 probability of being measured as 1.
Quantum gates alter qubit states, with measurements determining the outcome. Qubits remain in superposition until measured, at which point they collapse to 0 or 1 state.

Quantum computation also involves entanglement, connecting multiple qubits through multi-qubit quantum gates. 
These qubits cannot be treated as separate entities with individual probabilities of being 0 or 1. 
Instead, the measurement outcome of one qubit influences the measurement probabilities of others, a phenomenon known as entanglement. 
In systems with multiple entangled qubits, \begin{math}{2}^{n}\end{math} probability amplitudes are required to describe the quantum state of the $n$-qubit system.

\subsection{Quantum State Vector Simulation} \label{state}

The quantum state vector simulation is one of the general methods of quantum circuit simulation used in commercial quantum simulators \cite{quest, aleksandrowicz}.
Applying a quantum gate in the real quantum device corresponds to performing multiplication of the unitary matrix (of the gate) at the corresponding position of the qubit on the state vector.

Given a complex number of state vectors and gate-based quantum circuits for $n$-qubit system, the state vector simulation tracks full state vectors by applying each gate.
The state vector for the $n$-qubit quantum circuit has $2^{n}$ elements.
Thus, the memory requirement of the state vector for the $n$-qubit quantum circuit is a total of $2^{n+4}$ bytes, assuming that 16 bytes are required to describe the single complex number.
For example, a state vector describing 50 qubits has a memory requirement of 16 Petabytes, which exceeds the total main memory of Frontier, the world's highest-performing supercomputer \cite{bacou}.

\subsection{UPMEM PIM Architecture}

\begin{figure} [h] 
\centerline {
\includegraphics [width=\columnwidth] {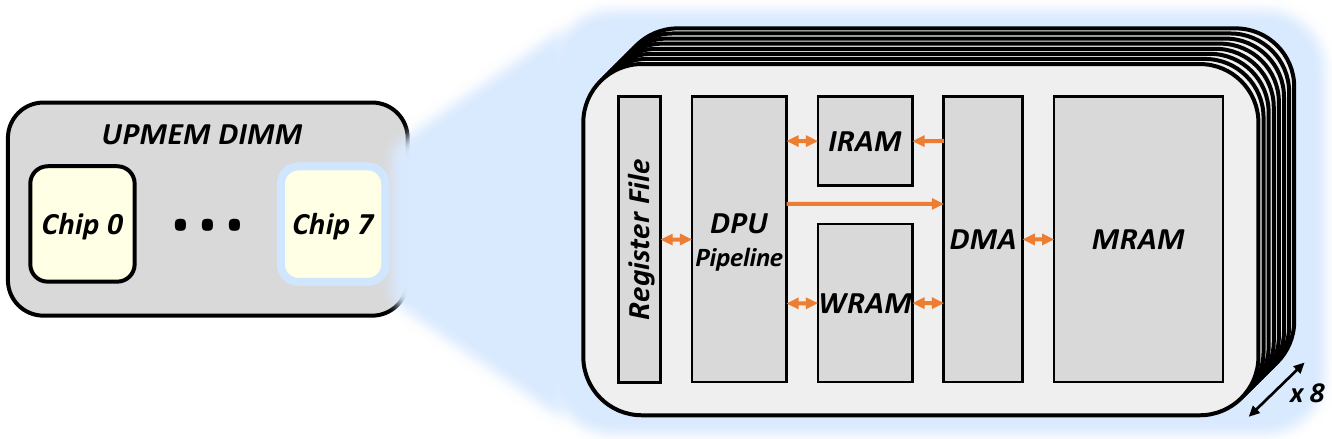} }
\caption {UPMEM PIM Organization
} 
  \Description[<short description>]{<long description>}
\label{UPMEM} 
\end{figure}

\noindent
Many PIM architectures were proposed to alleviate memory bandwidth limitations in various applications \cite{ahn2015scalable, gupta2019rapid, lee2021hardware, angizi, kal}. 
Despite these proposals, most works are implemented in simulators or simplified hardware prototypes, which makes it hard to realize the realistic constraints of the architecture. 
UPMEM, the first commercial PIM architecture, integrated processing elements called DPUs with DRAM chips. 
UPMEM PIM system can provide valuable insights to users and computer architects to help them understand the opportunities and limitations of PIM architecture. 

UPMEM DIMM includes eight memory chips in one rank, and each DIMM consists of two ranks. 
Each memory chip is coupled with eight DPUs. 
There are 20 UPMEM DIMMs in the server, thus providing 2,560 DPUs to the PIM system. 
\cref{UPMEM} shows an organization of eight UPMEM PIM chips. 
In a single chip, eight DPUs are connected to various elements.  
Each DPU is exclusively granted access to (i)  a 64MB slice of DRAM, \textit{Memory RAM (MRAM)}, (ii)  a 24KB instruction memory, \textit{Instruction RAM (IRAM)}, and (iii)  a 64KB scratchpad memory, \textit{Working RAM (WRAM)} \cite{gomez}. \textit{MRAM}s are available for data transfers between host CPUs and UPMEM DIMMs. 
Data for a DPU computation must be copied from \textit{MRAM} to \textit{WRAM} with the support of \textit{DMA} transfers. 
Considering the organization of a chip, programmers can exploit an additional resource of computation and a wider internal DIMM bandwidth as the number of UPMEM DIMMs increases.

\section{Motivation}

UPMEM PIM architecture offers several notable advantages. 
% First, it provides enhanced programmability through the UPMEM SDK, facilitating ease of use.
First, it provides the UPMEM SDK toolchain, including a C compiler and multiple function libraries, for the programmability of PIM architecture.
Second, UPMEM can provide significantly higher internal DRAM bandwidth for data processing, alleviating the constraints of the limited bandwidth between the CPU and memory \cite{devaux}.
% Despite these benefits, some limitations arise when applying quantum circuit simulation to UPMEM PIM.
Despite these benefits, there are several considerations to be taken into account when applying quantum circuit simulation to UPMEM PIM. 
Primarily, UPMEM PIM natively supports integer operations (such as addition and subtraction), with floating-point operations available solely via software emulation using library routines. 
Thus, the floating-point operations are far slower than the integer operations in UPMEM hardware. 
Additionally, direct inter-DPU communication is not supported, leading to expensive data transfers between the CPU and PIM memory for communication between DPUs \cite{gomez}. 
Therefore, it becomes essential to consider these implications in order to realize efficient quantum circuit simulations using the UPMEM hardware.

This section explains the workload characteristics of quantum circuit simulations and evaluates their performance using a roofline model analysis. 
It highlights the memory-intensive nature of these simulations, particularly in terms of data movement and matrix-vector multiplication operations, which often result in low locality of data access as the number of qubits increases. 
The section further discusses how these characteristics affect energy consumption and data access latency, arguing that Processing-In-Memory (PIM) architectures, such as UPMEM, may mitigate these challenges due to their broader internal DRAM bandwidth. 
Additionally, the section outlines the specific characteristics of UPMEM DIMMs, noting both the advantages and the implications, such as limited support for native floating-point operations and the lack of inter-DPU communication.

%UPMEM PIM offers several advantages stemming from the architecture. First, UPMEM provides programmability through the use of UPMEM SDK. Second, UPMEM represents broader internal DRAM bandwidth for processing data, rather than relying on the limited bandwidth between the CPU and memory \cite{devaux}. 
%Even though UPMEM PIM provides some benefits, there are some drawbacks when we adopt quantum circuit simulation to UPMEM PIM. First, UPMEM PIM supports only integer operations (e.g., addition, subtraction). Floating point operations are supported by software emulation through library routines. As shown in \cref{f3}, integer operations show average Z\% faster execution time in addition, subtraction, multiplication, and gemv operation, compared to floating point operations. Second, there is no support for inter-DPU communication. This results in costly data transfers between CPU and PIM memory to communicate with other DPUs \cite{gomez}.

\subsection{Workload Characteristics of Quantum Circuit Simulation}

\begin{figure} [h] 
\centerline {
\includegraphics [width=\columnwidth] {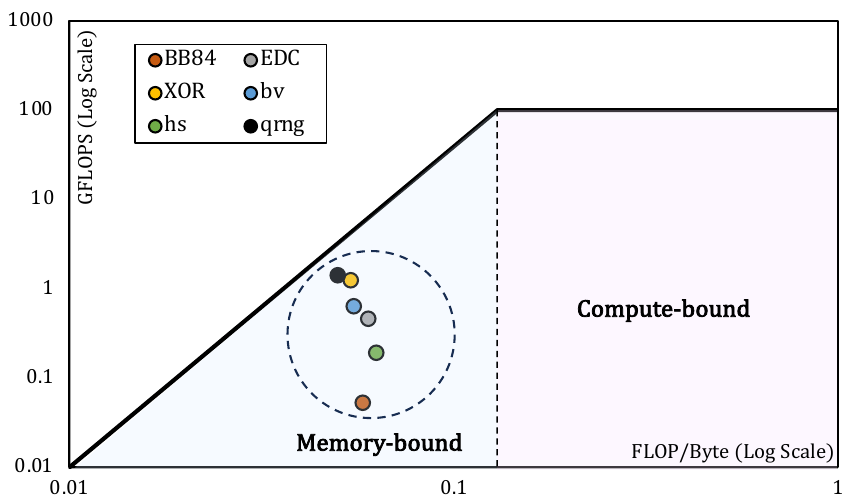} }
\caption {Roofline analysis of simulation workloads for benchmark quantum circuits.
CPU and benchmark information are detailed in \cref{t3} and \cref{t1} respectively.
} 
  \Description[<short description>]{<long description>}
\label{fm1} 
\end{figure}

\noindent 
Quantum circuit simulations require massive amounts of memory as the number of qubits increases, resulting in excessive data movement.
As described in \cref{state}, quantum circuit simulation requires $2^{n}$ state vector elements in ${n}$ qubit system.
Moreover, a substantial number of matrix-vector multiplication operations occur in quantum circuit simulations, frequently causing low locality data access to retrieve vector values. Particularly, the characteristic of low locality intensifies as the number of qubits increases, leading to a larger number of state vectors.
These properties of quantum circuit simulation have a detrimental influence on energy and data access latency. 
In \cref{fm1}, we collected roofline data for six benchmark quantum circuit simulation runs on CPU using Intel Advisor \cite{advisor}.
The roofline model analysis shows that these circuit simulation workloads are not performing optimally in the current architecture due to their memory-bounded operational characteristics. 
These analysis results suggest that adopting PIM for simulating quantum circuits can be efficient as we can take advantage of broader internal DRAM bandwidth.

%In \cref{fm1}, the roofline data was collected using Intel Advisor \cite{advisor} for simulation runs on 6 benchmark quantum circuits considered in this work. The roofline model analysis, as shown in \cref{fm1}, revealed that these simulation workloads are memory-bound, indicating their potential for efficient execution on PIM.

%\noindent We make a valuable finding that quantum circuits in simulation are memory-bound. Also, quantum circuit simulation requires massive amounts of memory, as the number of qubits increases, and this imposes excessive data movement. These properties have a detrimental influence on energy and latency. In this research, we propose quantum circuit simulation in PIM architecture to deal with the problems. We expect UPMEM PIM can be a promising solution to break the limitations of quantum circuit simulation with high internal bandwidth and low energy consumption.

%\subsection{Quantum Circuit Simulation in PIM}

\subsection{UPMEM DIMM Characteristic}

UPMEM architecture adopts the PIM concept, which exploits broader internal DRAM bandwidth for data processing and locates processing elements near memory. 
This can be an effective solution for memory-bound workloads, including quantum circuit simulation. 
However, we observe two implications to note for efficiently utilizing UPMEM to perform quantum circuit simulations.
First, DPU, the computational unit of UPMEM, natively supports integer addition, subtraction, and bit-wise operations \cite{gomez2}.
Floating-point or multiplication/division operations can be implemented using software library routines; however, they have lower throughput than natively supported operations \cite{gomez2}.
Second, there is no direct communication channel between DPUs \cite{lim}.
This can make it less scalable to handle workloads that require global communication.

\section{PIMutation Framework} \label{basic}

This section explains the organization and functionalities of the PI\textit{Mutation} framework, which is designed to facilitate quantum circuit simulation on the UPMEM PIM system. 
The framework consists of two major components: the host (CPU)-side, responsible for managing quantum state vectors and gates, and the DPU-side, which provides user primitives for performing quantum gate operations. 
Additionally, this section outlines three essential optimization methods implemented in PI\textit{Mutation}, which include gate merging to exploit native integer operations, row swapping to avoid matrix multiplication for certain quantum gates, and vector partitioning to handle separable quantum states. 
Each of these strategies is aimed at improving simulation efficiency by reducing computational complexity and communications between DPUs, considering the characteristics of UPMEM hardware.

\subsection{\textbf{Framework Overview}}

\begin{figure} [h] 
\centerline {
\includegraphics [width=1\columnwidth] {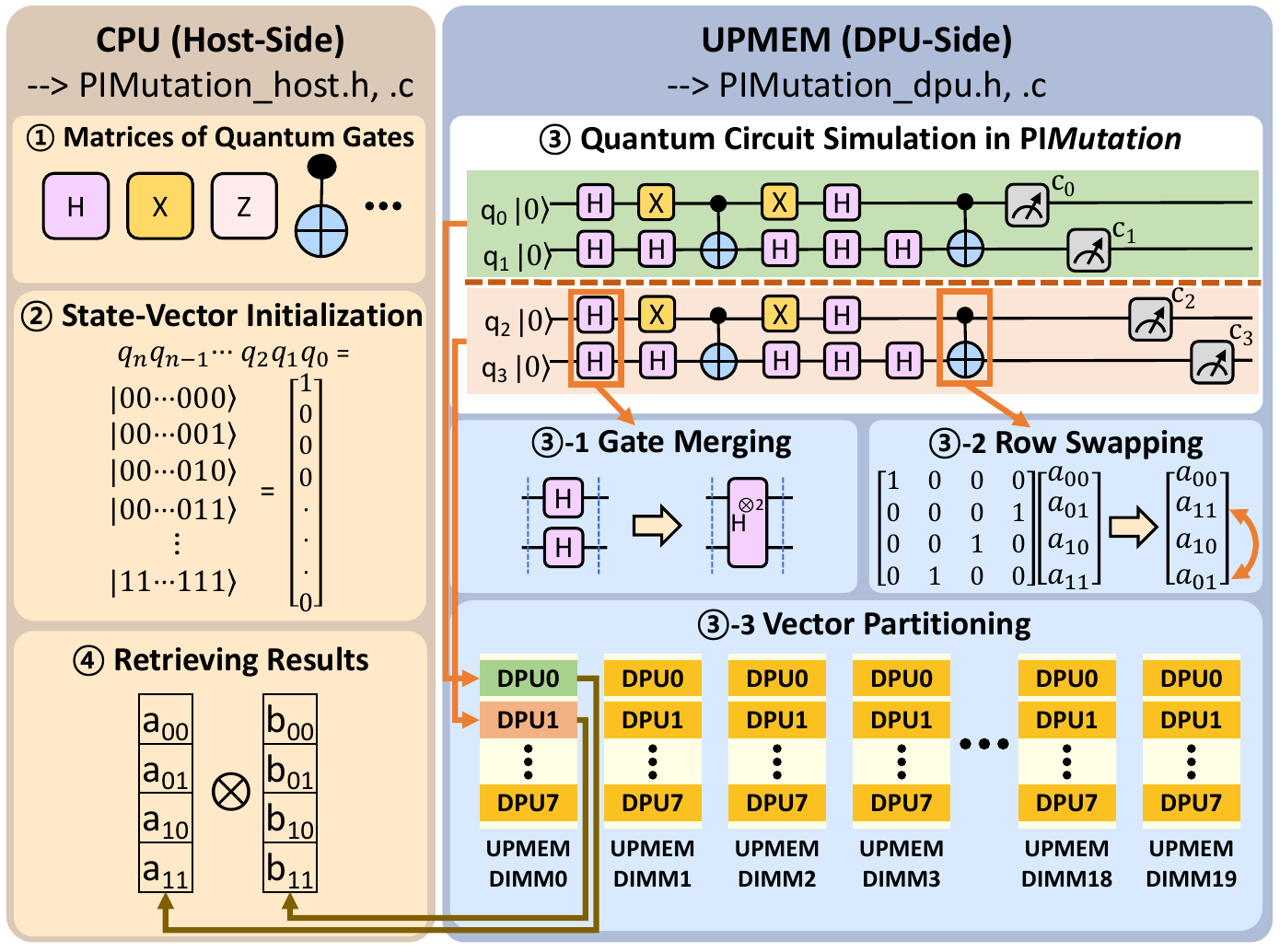} }
\caption {PI\textit{Mutation} framework overview}
  \Description[<short description>]{<long description>}
\label{f0} 
\end{figure}

\noindent
PI\textit{Mutation} framework takes into account the characterization of quantum circuit simulation and UPMEM PIM system. 
Our framework is implemented in C language and is mainly divided into host-side and UPMEM's DPU-side.

\textbf{Host-Side framework} supports fundamental functionalities for quantum circuit simulation. 
Quantum state vector simulation requires storing all quantum amplitudes before applying gates to circuits.
Moreover, quantum gates (represented by 2x2 or 4x4 matrices) are also needed for gate operation. 
These two significant functions are implemented in \texttt{PI\textit{Mutation}\_host.h} and \texttt{PI\textit{Mutation}\_host.c}, as shown in \cref{f0}. 

\textbf{UPMEM (DPU)-Side framework} provides user primitives for quantum circuit gate operation. 
The significant framework part, \texttt{PI\textit{Mutation}\_dpu.h} and \texttt{PI\textit{Mutation}\_dpu.c} as described in \cref{f0}, includes the basic operations between quantum gates and state vector. 
The framework part also provides three optimization methods described in \cref{GM}, \cref{rs}, and \cref{vp}.

% \textcolor{red}{The amplitudes involved in the gate operation depend on which qubit the quantum gate is located. 
% The algorithm for such state indexing is implemented in \texttt{PIANIST\_dpu.h}, and presented in Appendix A.1.}

%\noindent \textbf{Host-Side} framework supports fundamental functionalities for quantum circuit simulation. Quantum state vector simulation requires storing the whole quantum amplitudes before applying gates in circuits. Gates, represented by 2x2 or 4x4 matrices, are also needed for gate operation. Two significant functions are implemented in parisian\_host.h file as shown in \cref{f0}. In UPMEM SDK programming, the programmer decides the number of ranks, and DPUs to offload to UPMEM PIM. A programmer also should choose what data to offload to UPMEM considering precise data length. PARISIAN framework provides functions to facilitate PIM programmability, and ease the burden of programmers in other parts of the framework.

%\noindent \textbf{UPMEM (DPU)-Side} framework provides user primitives for quantum circuit gate operation. The significant framework part, parisian\_dpu.h as described in \cref{f0}, includes the basic operations between quantum gates and state vector. The framework part also provides three methods described later in \cref{Design}.
%The amplitudes involved in the gate operation depend on which qubit the quantum gate is located. The algorithm for such state indexing is implemented in parisian\_dpu.h, and presented in Appendix A.1.

% \section{PIANIST Design} \label{Design}

\subsection{\textbf{Gate Merging to Exploit Native Operations}} \label{GM}

\begin{figure} [h] 
\centerline {
\includegraphics [width=\columnwidth] {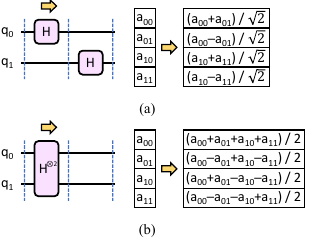} }
\caption {In case (a), the calculation of the Hadamard gate is performed individually, requiring $\sqrt2$ division operations. 
In contrast, in case (b), the implementation can be optimized by replacing the division with a shift operation by merging two Hadamard gates.} 
  \Description[<short description>]{<long description>}
\label{f2} 
\end{figure}

\noindent
Gate merging provides a way to avoid floating-point division operations by merging quantum gates and simulating quantum circuits with operations that UPMEM natively supports.
By eliminating floating-point operations, quantum gate operations can be computed by integer units, leading to faster quantum simulation results.
We observe that all elements of the matrices implementing RX or RY with rotation angles of $\pm\pi/2$ and $\pm3\pi/2$, and Hadamard gate have a magnitude of $1/\sqrt2$.
For example,

\begin{equation*}
\centering
H =
\begin{bmatrix}
    1 & 1 \\
    1 & -1
\end{bmatrix}/\sqrt{2}, \quad
RX(\pi/2) =
\begin{bmatrix}
    1 & -i \\
    -i & 1
\end{bmatrix}/\sqrt{2},
\end{equation*}

\begin{equation*}
\centering
RY(\pi/2) =
\begin{bmatrix}
    1 & -1 \\
    1 & 1
\end{bmatrix}/\sqrt{2}.
\end{equation*}
By merging them into even numbers, these gates can be implemented only by add, subtract, and shift operations (which is the native operation of UPMEM) by making the size of each element $1/2^n$ ($n$ is a non-negative integer).
\cref{f2} shows an example of avoiding division operations by merging the two Hadamard gates.

In addition, the quantum gates participating in the merge process do not necessarily have to be equivalent gates.
For example,

\begin{equation*}
\centering
H \otimes RY(\pi/2) =
\begin{bmatrix}
    1 & 1 \\
    1 & -1
\end{bmatrix}/\sqrt{2} \otimes
\begin{bmatrix}
    1 & -1 \\
    1 & 1
\end{bmatrix}/\sqrt{2}
\end{equation*}

\begin{equation*}
\centering
= \begin{bmatrix}
        1 & -1 & 1 & -1 \\
        1 & 1 & 1 & 1 \\
        1 & -1 & -1 & -1 \\
        1 & 1 & -1 & 1 \\
\end{bmatrix}/2.
\end{equation*}

\begin{figure} [h] 
\centerline {
\includegraphics [width=\columnwidth] {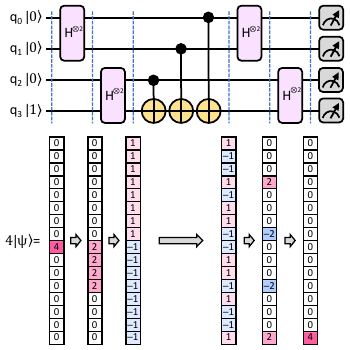} }
\caption {Integer-based 4-qubit BV algorithm circuit (the secret string $s$ = 111) simulation by using State quantization method.} 
  \Description[<short description>]{<long description>}
\label{f1} 
\end{figure}

Moreover, PI\textit{Mutation} simulates state amplitudes into integer types by examining the entire quantum circuit simulation process and then readjusting the starting amplitude value where possible.
For example, \cref{f1} multiplies the starting amplitude value by 4; thus, the amplitude of the entire simulation process is always an integer.
As a result, these strategies can treat both matrix gates and amplitudes as integer types, improving simulation speed.

\subsection{Row Swapping Instead of Matrix Multiplication} \label{rs}

We observe that some of the quantum gates are represented as a matrix where elements of each coefficient consist of only 0 or 1. Moreover, most of these gate operations serve to switch the state vectors of the corresponding qubits. Quantum gates such as
X, CNOT, Toffoli, and SWAP gates are representatives, which are frequently used in quantum circuits \cite{chuang}. Thus, simulations of these gates are implemented not by matrix vector multiplication but by simply swapping the corresponding row pairs within the state vector in PI\textit{Mutation}. The Row swapping optimization can be effective in the UPMEM PIM system because swapping state vectors are performed near DPUs.

\subsection{Vector Partitioning for Separable States} \label{vp}

\begin{figure} [h] 
\centerline {
\includegraphics [width=0.9\columnwidth] {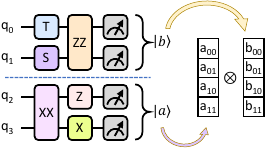} }
\caption {An example of a 4-qubit quantum circuit.
This quantum circuit can be separated into two 2-qubit separable quantum circuits.
The state vectors describing separated states ($\left|a\right>$, $\left|b\right>$) can be reconstructed to the full state vectors of the original 4-qubit quantum circuit through the tensor product.} 
  \Description[<short description>]{<long description>}
\label{f3} 
\end{figure}

\noindent 
This section describes how to separate and simulate quantum circuits into several sub-circuits that can be processed independently.
As shown in \cref{f3}, the qubit groups with no quantum gates applied to each other can be separated into individual qubit systems.
In this case, instead of simulating a 4-qubit system, two 2-qubit systems can be simulated independently.
Then, we can reconstruct the simulation results of the original quantum circuit by taking the tensor product to each state vector.
Moreover, depending on the application structure of the CNOTs, it can also be processed separately into properly independent sub-circuits \cite{jang}.

In this work, we designed these independent processing sub-circuits to be simulated in each UPMEM DPU and then reconstructed in the host CPU.
This significantly reduces memory requirements by reducing the simulated qubit scale and eliminates the need for data exchange between DPUs in the simulation process.
For example, assuming that 16 bytes are required to store a single complex number, separating a single 20-qubit quantum circuit into two 10-qubit quantum circuits significantly reduces the memory requirement of the required state vector from 16 MB to 32 KB.

\section{Experimental Methodology}

In this section, we describe our experiment configuration, execution scenario, benchmarks, and breakdown analysis of execution time. We utilize the UPMEM PIM system to validate the efficacy of the PI\textit{Mutation} framework for quantum circuit simulation. Furthermore, to more accurately validate the operation of PI\textit{Mutation}, we conduct the experiment by breaking down the execution time. 

\subsection{Configuration and Execution Scenario}
\noindent\textbf{\textit{Configuration:}} 
Specifications and configurations of the UPMEM server used in our experiments are given in \cref{t3}.
% We perform the energy comparison between UPMEM and CPU by comparing the energy consumption of the PIM system and CPU \cite{gomez2}. 
In the UPMEM server system we use for evaluation, the DRAM without the PIM module has a capacity of 256 GB, which can perform full state vector simulations of up to 34 qubits without applying the techniques or schemes proposed in this work.
Considering the UPMEM PIM system, where data is transferred from the main memory to the PIM module, full state vector simulation using the PI\textit{Mutation} framework in the UPMEM PIM server can handle up to 33 qubits.

% \begin{figure*} [h]
% \centerline {
% \includegraphics [width=2.0\columnwidth] {fig/single_figure.pdf} }
% \caption {Comparison of simulation execution time and relative speedup by benchmark circuit according to the application of the optimization techniques. The speedup is derived from comparing \textbf{Baseline} to \textbf{GM} and \textbf{RS} measured in UPMEM.} 
% \label{single_figure} 
% \end{figure*}

\begin{table}[h]
\renewcommand{\arraystretch}{1}
\renewcommand{\tabcolsep}{0.3mm}
\caption{The Platform for Experiments}
\label{t3}
\centering
%\small
\begin{tabular}{|c|c|}
\hline
CPU Model &  Intel(R) Xeon(R) Silver 4215 \\
\hline
Sockets & 2 \\
\hline
% CPU MHz & 1,000 \\
% \hline
\multirow{2}{*}{Memory} & 4 x 64 GB DDR4-2666 \\
 & RDIM Dual Rank DRAM (256 GB) \\
\hline
PIM Memory & 20 $\times$ DDR4-2400 PIM Modules (160 GB) \\
\hline
UPMEM SDK version & 2023.2.0 \\
\hline
QuEST version & v3.7.0 \cite{quest} \\
\hline
% \multirow{2}{*}
{Compiler version} & gcc 8.3.0 \\
% & Python 3.7.3 \\
\hline
{CMake version} & 3.13.4 \\
\hline
Energy Measurement & Intel RAPL \cite{RAPL} \\
\hline
\end{tabular}
\end{table}

\noindent\textbf{\textit{Execution Scenario:}} We evaluate PI\textit{Mutation} framework using four different versions of a simulation for Single-DPU evaluation:
\begin{itemize} 
\item\textbf{Baseline}: Baseline implements a na\"ive full-state vector operation based on matrix-vector multiplication using UPMEM DPU. 

\item \textbf{GM}: As discussed in \cref{GM}, the first optimization technique of PI\textit{Mutation}, Gate merging, is implemented.

\item \textbf{RS}: As discussed in \cref{rs}, Row Swapping technique is implemented.

\item \textbf{GM+RS}: In this version, both \textbf{GM} and \textbf{RS} techniques is applied.

\item \textbf{VP}: As discussed in \cref{vp}, Vector partitioning is implemented.
% We consider VP on the cases of the multiple-DPU evaluation.}
\end{itemize}

% \noindent\textbf{\textit{DPU Programming:}} The DMA engine can copy up to 2 KB with a single command, and in this task, we send a state vector of 2 KB per DPU \cite{nider}.
% In this work, the state amplitude is considered as 4 Byte Integer or 8 Byte Double type, so the qubit size of the full state vector managed by a single DPU is up to 8 qubits.

\subsection{Benchmarks and Breakdown Analysis}
\noindent\textbf{\textit{Benchamarks:}} The benchmark quantum circuits used in these experiments are presented in \cref{t1}.
The scale of the benchmark circuit depends on the value of $n$.
\begin{table}[h]
\centering
%\small
\renewcommand{\arraystretch}{1}
\renewcommand{\tabcolsep}{0.7mm}
\caption{
Evaluated Benchmark Quantum circuits}
\label{t1}
\begin{tabular}{|c|c|c|c|c|c|}
\hline
  \multirow{2}{*}{Algorithm \big[Reference\big]} & \multirow{2}{*}{Qubits} & \# of 1Q & \# of 2Q &  \multirow{2}{*}{Abbr.} \\
 &  & Gates & Gates & \\
\hline
\hline
BB84 Protocol \cite{bennett} & n & 2n & 0 & BB\_n \\
\hline
Bernstein--Vazirani \cite{bv} & n & 2n & n-1 & BV\_n \\
\hline
Error Detection Code \cite{devitt} & n & 2n & 2n-2 & EDC\_n \\
 \hline
Hidden Subgroup & \multirow{2}{*}{2n} & \multirow{2}{*}{6n} & \multirow{2}{*}{2n} & \multirow{2}{*}{HS\_2n} \\
Problems \cite{li3} & & & & \\
 \hline
Quantum Random & \multirow{2}{*}{n} & \multirow{2}{*}{n} & \multirow{2}{*}{0} & \multirow{2}{*}{QRNG\_n} \\
Number Generator \cite{li3} & & & & \\
 \hline
Exclusive-OR \cite{revlib} & n & 0 & n-1 & XOR\_n \\
\hline
\end{tabular}
\end{table}

\noindent\textbf{\textit{Breakdown Analysis of Execution Time:}} The simulation run-times discussed in \cref{results} are broken down into the following components and analyzed.

\begin{itemize}
\item\textbf{C-to-D Tran.}: Time to transfer data from the CPU to the DPU.
The data transmitted includes initial state vector information and quantum gates (matrix) required by the circuit.

\item \textbf{Comp.}: Time the computation takes place inside the DPU.

\item \textbf{D-to-C Tran.}: Time to transfer data from the DPU to the CPU.
Returns the result state vector processed by the DPU back to the host CPU.

\item \textbf{Recon.}:
Required reconstruction time if split state vectors are sent to multiple DPUs.
As discussed in \cref{vp}, it is the overhead of reconfiguring the original state vector from the host CPU after processing from the DPU.
\end{itemize}

% \begin{figure*} [h]
% \centerline {
% \includegraphics [width=1.9\columnwidth] {fig/single_dpu.pdf} }
% \caption {Comparison of simulation execution time and relative speedup by benchmark circuit according to the application of the optimization techniques.} 
% \label{single} 
% \end{figure*}

\section{Results and Analysis} \label{results}

This section explains the performance analysis of PI\textit{Mutation} under both single-DPU and multi-DPU configurations. 
It delves into the specific optimizations achieved through Gate Merging, Row Swapping, and Vector Partitioning techniques, highlighting their impact on execution time and energy efficiency.
Additionally, it compares the results of PI\textit{Mutation} against the QuEST simulator on a CPU, providing insights into the scalability and effectiveness of the proposed framework in handling quantum circuit simulations with varying qubit benchmarks.

\subsection{Single-DPU Performance} \label{single_perf}

\begin{figure} [h] 
\centerline {
\includegraphics [width=\columnwidth] {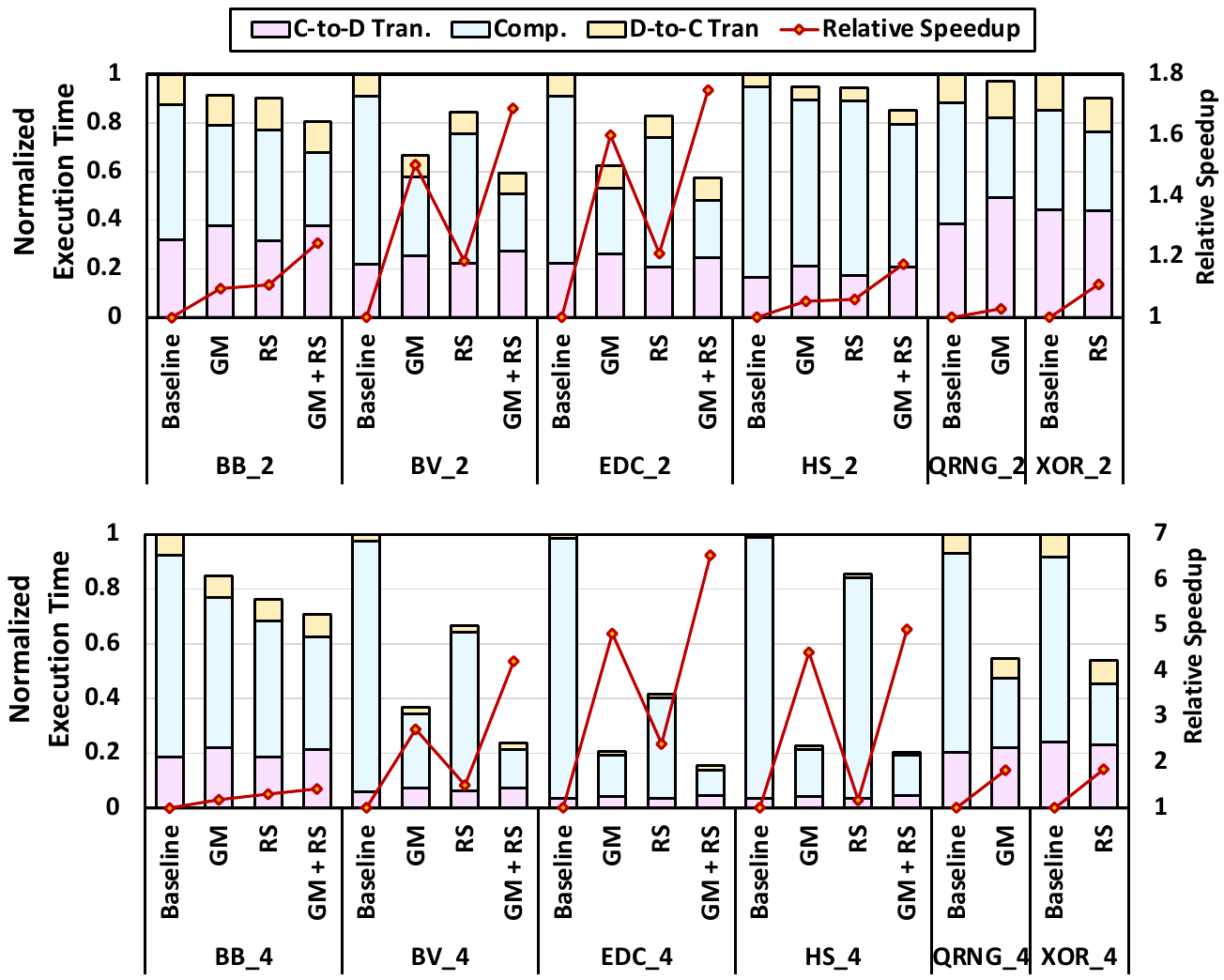} }
\caption {Comparison of the normalized execution time and relative speedup according to the application of the optimization techniques in two- and four-qubit benchmark circuits.} 
  \Description[<short description>]{<long description>}
\label{single_figure} 
\end{figure}

\noindent
In this section, we evaluate PI\textit{Mutation} by dividing the experiment depending on the number of DPUs. 
In the single-DPU experiment, we analyzed the performance of gate merging and row swapping by breaking down the execution time across six benchmarks. 
In the multi-DPU experiment, we evaluate the speedup and energy consumption of PI\textit{Mutation} by comparing it with the QuEST simulator on CPU \cite{quest}. 
Given that PI\textit{Mutation} represents the first research for quantum circuit simulation using commercialized PIM, it is hard to compare with other PIM frameworks. 
Additionally, as the UPMEM PIM system does not support GPUs, comparing the PI\textit{Mutation} with GPU-based studies \cite{zhao2, gutierrez2008parallel} presents challenges. 
Thus, we compare PI\textit{Mutation} with the QuEST simulator, which is written in C language and compatible with the UPMEM system.

\cref{single_figure} presents the normalized execution time of PI\textit{Mutation} framework in a single-DPU experiment. 
In the two-qubit benchmarks, \textit{Gate Merging} technique, \texttt{GM}, achieved a speedup of 1.26$\times$, while the \textit{Row Swapping} technique, \texttt{RS}, resulted in a 1.13$\times$ speedup compared to \texttt{Baseline}. 
In the four-qubit benchmarks, the speedups are 2.99$\times$ for \texttt{GM} and 1.65$\times$ for \texttt{RS}. In the QRNG\_n benchmark, \texttt{RS} cannot be applied, and in the XOR\_n benchmark, \texttt{GM} technique is not applicable, resulting in the exclusion of execution time and speedup data from \cref{single_figure}.  
When both \textit{Gate Merging} and \textit{Row Swapping} techniques, \texttt{GM+RS}, were employed, a speedup of 1.46$\times$ was achieved in the two-qubit benchmarks, and a 4.27$\times$ speedup was observed in the four-qubit benchmarks relative to the Baseline.
As the number of qubits processed by a single-DPU increases, the speedup from applying \texttt{GM} and \texttt{RS} shows an increase.

Upon breaking down the execution time for the two-qubit benchmarks, \textit{Time required to transfer data from the CPU to the DPU}, referred to as \texttt{C-to-D Tran.}, accounted for an average of 38.78\% in the \texttt{GM} technique and 30.49\% in the \texttt{RS} technique.
In terms of \textit{computation Time}, referred to as \texttt{Comp.}, it constituted 48.57\% of the overall execution time for \texttt{GM} and 58.04\% for \texttt{RS} technique.
\textit{Time required to transfer data from the DPU back to the CPU}, denoted as \texttt{D-to-C Tran.}, accounted for 12.65\% of the total execution time in \texttt{GM} technique and 11.47\% in the \texttt{RS} technique.
In the four-qubit benchmarks, \texttt{Comp.} significantly increases, comprising 66\% for \texttt{GM} technique and 75\% for \texttt{RS} technique of the total execution time.

% \textbf{GM} reduces the execution time difference by 11.7$\times$, 16.6$\times$, and 91.8$\times$. The execution time gap can be narrowed by 11.5$\times$, 31.7$\times$, and 265.3$\times$ when \textbf{RS} technique is implemented. 
% As \textbf{GM+RS} is deployed in UPMEM, it can minimize the time difference by 11.4$\times$, 14$\times$, 90$\times$ compared to CPU. Among two optimizations in the single-DPU performance analysis, \textbf{GM} achieves more enhancement regarding execution time. 
% \texttt{GM+RS} improves simulation time by 86.2\% and reduces energy consumption by 14.5\% compared to \texttt{Baseline}.
% Note that \texttt{GM} contributes more to performance improvement for workloads with many single-qubit gates and \texttt{RS} for workloads with many 2-qubit gates (e.g., CNOT).

% \cref{single_figure} also shows that speedup is enhanced when two optimization schemes are implemented in higher qubit benchmarks. EDC\_n benchmark demonstrates significant improvement in speedup, because \textbf{GM} can be frequently applied in the quantum circuit. On the other hand, BB\_n benchmark exhibits a relatively low speedup, as a small number of \textbf{GM} is implemented.

\begin{figure} [h] 
\centerline {
\includegraphics [width=0.92\columnwidth] {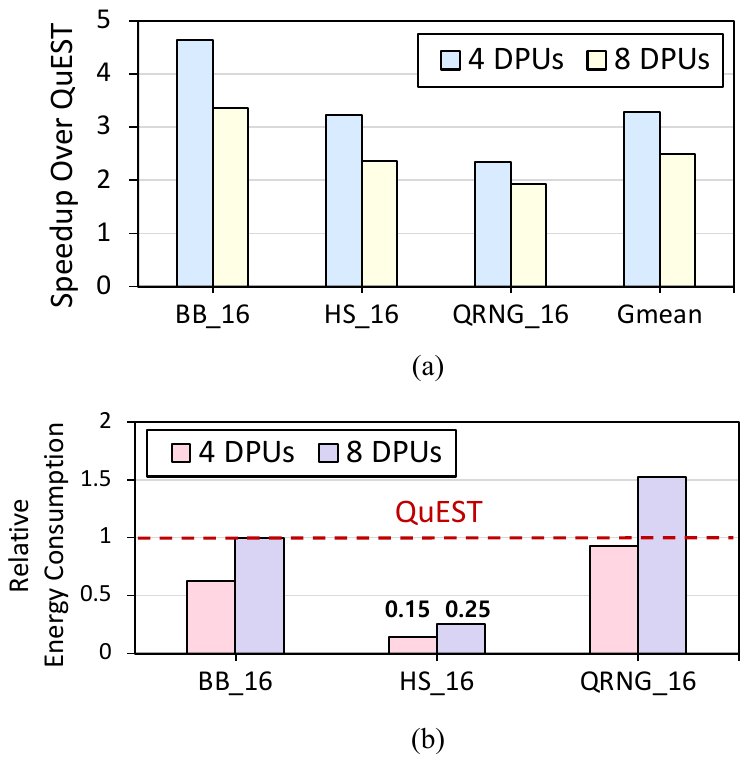} }
\caption {(a) shows a speedup of PI\textit{Mutation} over QuEST simulator on CPU in 16-qubit benchmarks. Speedup is derived by comparing the execution time of PI\textit{Mutation} with the QuEST simulator executed on the CPU.
(b) shows relative energy consumption of PI\textit{Mutation} compared to QuEST simulator.} 
  \Description[<short description>]{<long description>}
\label{16Q} 
\end{figure}

\subsection{Multi-DPU Performances by Partitioning}

% \begin{figure} [h] 
% \centerline {
% \includegraphics [width=0.9\columnwidth] {fig/32Q.pdf} }
% \caption {(a) shows simulation time decomposition analysis and total time comparison by applying vector partitioning techniques.
% (b) shows a simulation time comparison of PIANIST using multiple DPUs normalized to the QuEST simulator run-time.} 
% \label{32Q} 
% \end{figure}
% \cref{16Q} and \ref{32Q} shows the performance of PI\textit{Mutation} by applying vector partitioning (\textbf{VP}) techniques in 16- and 32-qubit benchmarks. In the multi-DPU experiment, we compare PI\textit{Mutation} with CPU-based QuEST simulator. We evaluate the performance of PI\textit{Mutation} and QuEST simulator by conducting tests on 16- and 32-qubit benchmarks.
% Note that PIANIST sends each sub-state vector divided by the vector partitioning technique to each DPU.

Experiment results for 16-qubit benchmarks, which incorporate \texttt{GM}, \texttt{RS}, and \textit{Vector Partitioning}, \texttt{VP} techniques, are represented in \cref{16Q}.
In the Multi-DPU experiment, we conducted tests on the BB\_n, HS\_n, and QRNG\_n benchmarks, which are applicable to \texttt{VP} techniques.
PI\textit{Mutation} achieves an average 3.42$\times$ and 2.56$\times$ speedup over the QuEST simulator in 4 and 8 DPUs, respectively, shown in \cref{16Q} (a).
As shown in \cref{16Q} (a), configurations with 4 DPUs, each processing four qubits, achieve a higher speedup compared to configurations with 8 DPUs, each processing two qubits.
In this multi-DPU experiment, we observe that allocating an optimal number of qubits per single-DPU reduces the times for \texttt{C-to-D Tran.}, \texttt{D-to-C Tran.}, and Reconstruction time in \texttt{VP}, referred to as \texttt{Recon.}, positively influencing the speedup of execution time.
From an energy consumption perspective in \cref{16Q} (b), PI\textit{Mutation} demonstrated an average energy consumption of 56.79\% and 92.76\%, respectively, when utilizing 4 and 8 DPUs, compared to the QuEST simulator.

\begin{figure} [h] 
\centerline {
\includegraphics [width=0.92\columnwidth] {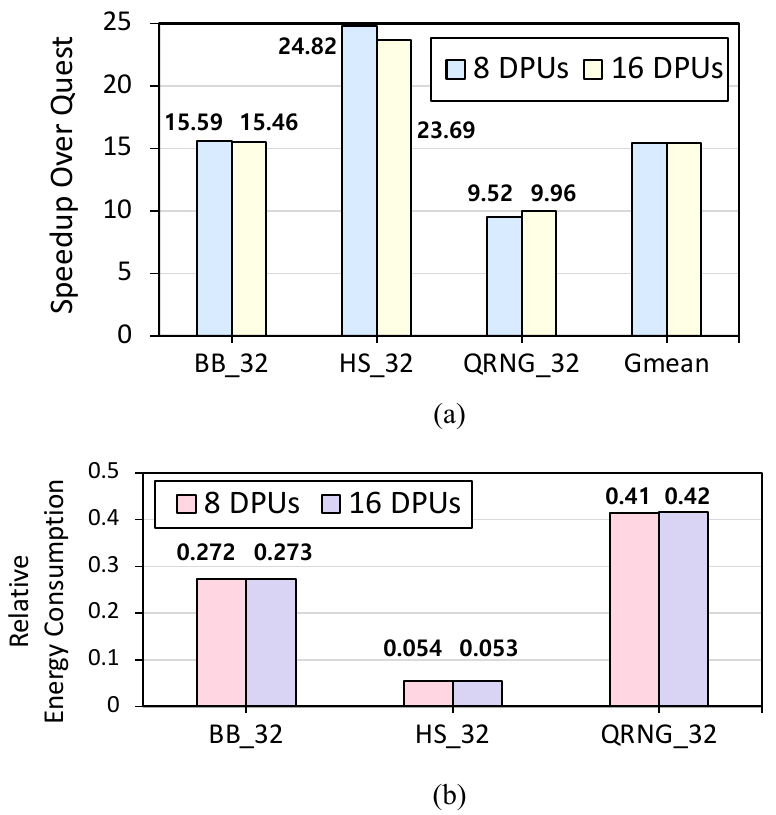} }
\caption {Speedup and relative energy consumption of PI\textit{Mutation} over QuEST simulator executed on CPU in 32-qubit benchmarks.} 
  \Description[<short description>]{<long description>}
\label{32Q} 
\end{figure}

\cref{32Q} presents the results for the 32 qubit benchmarks, which require a total of 64GB to store the state vectors.
PI\textit{Mutation} demonstrates an average 16.64$\times$, 16.37$\times$ speedup for all benchmarks in 8 DPUs and 16 DPUs over the QuEST simulator. 
Compared to the 16-qubit benchmarks, 32-qubit benchmarks exhibit greater improvements in terms of speedup.
The reason for the improvements is that PI\textit{Mutation} leverages the advantage of PIM to minimize data transfers, while the QuEST simulator suffers from excessive data transfer between CPU and main memory.
Energy consumption also decreased to 24.71\% when utilizing 8 DPUs and 16 DPUs compared to the QuEST simulator. The reduction in energy consumption is also attributed to the decreased data transfers in the PI\textit{Mutation} framework.
As shown in \cref{32Q}, both speedup and relative energy consumption exhibit similar results at configurations of 8 DPUs and 16 DPUs.
This is attributed to the \texttt{Recon.} process, which retrieves results from the PIM to CPU in the \texttt{VP} technique and accounts for over 80\% of the total execution time.

\section{Related Work}
Various techniques have been proposed to divide quantum circuits for efficiency, such as the proposed vector partitioning.
These techniques include quantum circuit cutting techniques that include wire-cutting \cite{kan2024scalable} or gate-cutting \cite{ren2024hardware}.
Compared to these techniques, the proposed vector partitioning has the advantage of less circuit reconstruction overhead instead of being applicable only to separable states.

\section{Conclusion}
In this work, we propose PI\textit{Mutation}, a quantum circuit simulation framework that leverages UPMEM PIM architecture. 
For the first time, we implement the PIM framework for quantum circuit simulation using a real PIM system.
We propose three efficient simulation techniques that consider both the computational characteristics of quantum circuits and the hardware attributes of UPMEM DIMMs.
Proposed strategies demonstrate significant improvements in simulation speed and reductions in energy consumption compared to the QuEST simulator on CPU. 
While the design presented in this work may not be a flawless solution, it underscores the potential of future PIM technologies and can open up new avenues for quantum circuit simulation research.
% We explored the potential of quantum circuit simulation efficiency in the first commercial PIM hardware, and hope that this exploration of possibilities will be broadly extended and compared to other near-data processing hardware or emerging architectures in the future.
We hope that this exploration of possibilities will be broadly extended in the future.

\begin{acks}
The authors extend their appreciation to reviewers who have provided insightful feedback to improve the quality of this paper.
%Additionally, this work was partially supported by another IITP grant funded by the Korean government (MSIT) (No. 2024-00402898, Simulation-Based High-Speed/High-Accuracy Data Center Workload/System Analysis Platform).
This research was partially funded by the National Research Foundation of Korea (NRF), supported by the Korean government (Ministry of Science and ICT (MSIT)) under the project Creation of the Quantum Information Science R\&D Ecosystem based on Human Resources (No. RS-2023-00303229).
Additionally, this research was partially funded by a grant from the Institute of Information \& Communications Technology Planning \& Evaluation (IITP), supported by the Korean government (Ministry of Science and ICT, MSIT) (No. 2024-0-00441, Memory-Centric Architecture Using Reconfigurable PIM Devices). 
Won Woo Ro is the corresponding author, and the correspondence regarding this work should be directed to him.
\end{acks}

\bibliographystyle{ACM-Reference-Format}
\bibliography{sample-base}

\end{document}